\documentclass[10pt]{article}
\usepackage{graphicx}
\usepackage{amsmath}
\usepackage{amssymb}
\usepackage{caption2}
\begin{document}
\bibliographystyle{prsty}
\begin{center}
{\large {\bf \sc{ Doubly-Charm and Doubly-Bottom Pentaquark molecular States via the QCD sum rules }}} \\[2mm]
Xiu-Wu Wang, Zhi-Gang  Wang\footnote{E-mail: zgwang@aliyun.com.  }\\
 Department of Physics, North China Electric Power University, Baoding 071003, P. R. China
\end{center}

\begin{abstract}
In the present work, the doubly-charm  and doubly-bottom pentaquark molecular states $D^{(*)}\Sigma_c^{(*)}$ and $B^{(*)}\Sigma_b^{(*)}$ are studied via the QCD sum rules. Sixteen color singlet-singlet type  currents with the definite isospin-spin-parity $IJ^P$ are constructed to interpolate  the corresponding  hadronic states with the same quantum numbers.  The masses and pole residues of those doubly-heavy pentaquark molecular  states are calculated, the results show that their masses are all below the corresponding  meson-baryon thresholds, which means that they are possible bound states, not resonant states,  moreover, the possible decay channels for the doubly-charm molecular states are given.
\end{abstract}



\section{Introduction}
Since the discovery of the exotic state $X(3872)$ by the Belle collaboration \cite{Bell-3872}, many charmonium-like and bottomonium-like states have been observed by different experimental groups \cite{PDG}. Due to the extraordinary properties of those exotic $X/Y/Z$ states, it is hard to embed them into the conventional quarkonium spectroscopy, revealing their natures becomes into an important and hot branch of hadron physics. The existing interpretations of their nature  contain the multi-quarks, hybrid, glueball, etc. The interesting
phenomenon is that many of them have the masses near the thresholds of the meson-meson or meson-baryon pairs, so it is natural to interpret them as the hadronic molecules consist of the color-singlet clusters  \cite{GFK-review-hadronic-molecule}.

In 2015,  the LHCb collaboration observed two pentaquark candidates $P_c(4380)$ and $P_c(4450)$ via the analysis of the decay $\Lambda_b^0\rightarrow J/\psi K^-p$ \cite{LHCb-Pentaquark-4450}. In 2019, its improved experiment showed that the $P_c(4450)$ consists  of two narrow overlapping peaks $P_c(4440)$ and $P_c(4457)$, moreover, the new pentaquark candidate $P_c(4312)$ was reported \cite{LHCb-Pentaquark-4312}. Many theoretic groups proposed that those exotic $P_c$ states are the hidden-charm  pentaquark molecular states $\bar{D}^{(*)}\Sigma_c^{(*)}$ \cite{Pentaquark-Molecule-ChenR,Pentaquark-Molecule-ChenHX,Pentaquark-Molecule-Roca,
 Pentaquark-Molecule-HeJ,Pentaquark-Molecule-LiuMZ,Pentaquark-Molecule-XiaoCW,Pentaquark-Molecule-WangZG,
 Pentaquark-Molecule-WangZG,XWWang-penta-mole}.
Since the spin-parity $J^P$ of those $P_c$ states have not been determined  experimentally yet, their physical natures are still under hot debates.

In 2017, the doubly-charmed baryon $\Xi_{cc}^{++}$ was observed by LHCb collaboration in the $\Lambda_c^+K^-\pi^+\pi^+$ mass distribution \cite{LHCb-Xicc-observed}.  In 2021, the  doubly-charm tetraquark candidate $T_{cc}^+$ was observed by the LHCb collaboration via the analysis of the $D^0D^0\pi$ mass spectrum \cite{LHCb-Tcc-observed,LHCb-Tcc-observed-2}.
They show evidences  that two charm quarks can correlate in a hadronic state, and has trigged high enthusiasms  of different theoretic groups to study the possible doubly-charm (doubly-bottom) tetraquark states $T_{cc}$ ($T_{bb}$). In fact, various theoretical  methods have  been applied to study them, such as the  color-magnetic interaction model \cite{XLiu-Tcc-2017,Cheng-Tcc-2021}, (simple) potential  quark models \cite{Karliner-Tcc-2017,GYang-Tcc-2020,HWKe-Tcc-BS}, QCD sum rules \cite{WZG-APPB,ZGW-Tcc-2018,CFQiao-Tcc-2020,Agaev-Tcc-decay}, lattice QCD simulations \cite{Francis-Tcc-2019,Junnarkar-Tcc-2019,Francis-Tcc} and so on \cite{QQin-Tetraquark,UOzdem-Tcc-magnet,Estia-Tcc}.

All in all, the hidden-charm pentaquark (molecular) states $P_c$, doubly-charm baryon $\Xi_{cc}^{++}$, doubly-charm tetraquark (molecular) state $T_{cc}^+$ do exist, it is natural to ask the question: could the doubly-charm (doubly-bottom) pentaquark (molecular) states  exist? There have been several works on this subject \cite{Penta-mole-cc-ChenR,Penta-mole-cc-DongXK,Penta-mole-cc-ChenDY,Xliu-DB-penta-2021,
 ZGWang-Xicc,Penta-mole-cc-YangG,Penta-mole-cc-Shimizu,Penta-mole-cc-GuoZH,Penta-mole-cc-ChenW}.

In Ref.\cite{ZGWang-Xicc}, we studied the doubly-heavy baryon states and pentaquark states with the QCD sum rules. In Ref.\cite{XWWang-penta-mole}, we applied the isospin eigenstate currents to study the hidden-charm pentaquark molecular states via the QCD sum rules for the first time, and argued that the $P_c(4312)$, $P_c(4380)$, $P_c(4440)$ and $P_c(4457)$ can be assigned as the hadronic molecular states with the isospin $I=\frac{1}{2}$, the lower isospin eigenstates. In Ref.\cite{XWWang-pentawiths}, the hidden-charm pentaquark molecular states with a valence $s$ quark are studied, our results show that the lower  isospin eigenstates may form the hadronic molecular states. Inspired by all those studies, the doubly-heavy pentaquark molecular states are studied in the framework of QCD sum rules in the present work.

The article is arranged as follows: the QCD sum rules for the pentaquark molecular states are obtained in Sect.2; the numerical results and discussions are given  in Sect.3; Sect.4 is reserved for the conclusions.

\section{QCD sum rules for the pentaquark molecular states}
In the present work, the following color-singlet currents are chosen to interpolate the corresponding mesons and baryons with the same quantum numbers,
\begin{eqnarray}
J^{D^0}(x)&=&\bar{u}(x)\textsf{i}\gamma_5c(x)\, ,\nonumber \\
J^{D^+}(x)&=&-\bar{d}(x)\textsf{i}\gamma_5c(x)\, ,\nonumber \\
J^{D^{*0}}_{\mu}(x)&=&\bar{u}(x)\gamma_{\mu}c(x)\, ,\nonumber \\
J^{D^{*+}}_{\mu}(x)&=&-\bar{d}(x)\gamma_{\mu}c(x)\, ,\nonumber \\
J^{\Sigma_c^+}(x)&=&\varepsilon^{ijk}u^{i\texttt{T}}(x)C\gamma_{\mu}d^j(x)\gamma^{\mu}\gamma_5c^k(x)\, ,\nonumber\\
J^{\Sigma_c^{++}}(x)&=&\varepsilon^{ijk}u^{i\texttt{T}}(x)C\gamma_{\mu}u^j(x)\gamma^{\mu}\gamma_5c^k(x)\, , \nonumber\\
J^{\Sigma_c^{*+}}_{\mu}(x)&=&\varepsilon^{ijk}u^{i\texttt{T}}(x)C\gamma_{\mu}d^j(x)c^k(x)\, , \nonumber\\
J^{\Sigma_c^{*++}}_{\mu}(x)&=&\varepsilon^{ijk}u^{i\texttt{T}}(x)C\gamma_{\mu}u^j(x)c^k(x)\, ,
\end{eqnarray}
\begin{eqnarray}
J^{B^-}(x)&=&\bar{u}(x)\textsf{i}\gamma_5b(x)\, ,\nonumber \\
J^{B^0}(x)&=&-\bar{d}(x)\textsf{i}\gamma_5b(x)\, ,\nonumber \\
J^{B^{*-}}_{\mu}(x)&=&\bar{u}(x)\gamma_{\mu}b(x)\, ,\nonumber \\
J^{B^{*0}}_{\mu}(x)&=&-\bar{d}(x)\gamma_{\mu}b(x)\, ,\nonumber \\
J^{\Sigma_b^0}(x)&=&\varepsilon^{ijk}u^{i\texttt{T}}(x)C\gamma_{\mu}d^j(x)\gamma^{\mu}\gamma_5b^k(x)\, ,\nonumber\\
J^{\Sigma_b^{+}}(x)&=&\varepsilon^{ijk}u^{i\texttt{T}}(x)C\gamma_{\mu}u^j(x)\gamma^{\mu}\gamma_5b^k(x)\, , \nonumber\\
J^{\Sigma_b^{*0}}_{\mu}(x)&=&\varepsilon^{ijk}u^{i\texttt{T}}(x)C\gamma_{\mu}d^j(x)b^k(x)\, , \nonumber\\
J^{\Sigma_b^{*+}}_{\mu}(x)&=&\varepsilon^{ijk}u^{i\texttt{T}}(x)C\gamma_{\mu}u^j(x)b^k(x)\, ,
\end{eqnarray}
where $\textsf{i}^2=-1$, the $i$, $j$ and $k$ are the color indexes, the $C$ represents the charge conjugation matrix, the superscript $\texttt{T}$ means the transpose of the quark fields in the Dirac spinor space. All the currents are isospin eigenstates, accordingly, the meson-baryon type pentaquark currents of isospin eigenstates are constructed as,
\begin{eqnarray}
J_{\frac{1}{2}}^{D\Sigma_c}(x)&=&\frac{1}{\sqrt{3}}J^{D^+}(x)J^{\Sigma_c^+}(x)-\sqrt{\frac{2}{3}}J^{D^0}(x)J^{\Sigma_c^{++}}(x) \, , \nonumber\\
J_{\frac{3}{2}}^{D\Sigma_c}(x)&=&\sqrt{\frac{2}{3}}J^{D^+}(x)J^{\Sigma_c^+}(x)+\frac{1}{\sqrt{3}}J^{D^0}(x)J^{\Sigma_c^{++}}(x)\, ,\nonumber\\
J_{\frac{1}{2};\mu}^{D\Sigma_c^*}(x)&=&\frac{1}{\sqrt{3}}J^{D^+}(x)J^{\Sigma_c^{*+}}_{\mu}(x)-\sqrt{\frac{2}{3}}J^{D^0}(x)J^{\Sigma_c^{*++}}_{\mu}(x)\, ,\nonumber\\
J_{\frac{3}{2};\mu}^{D\Sigma_c^*}(x)&=&\sqrt{\frac{2}{3}}J^{D^+}(x)J^{\Sigma_c^{*+}}_{\mu}(x)+\frac{1}{\sqrt{3}}J^{D^0}(x)J^{\Sigma_c^{*++}}_{\mu}(x)\, ,\nonumber\\
J_{\frac{1}{2};\mu}^{D^{*}\Sigma_c}(x)&=&\frac{1}{\sqrt{3}}J^{D^{*+}}_{\mu}(x)J^{\Sigma_c^+}(x)-\sqrt{\frac{2}{3}}J^{D^{*0}}_{\mu}(x)J^{\Sigma_c^{++}}(x)\, ,\nonumber\\
J_{\frac{3}{2};\mu}^{D^{*}\Sigma_c}(x)&=&=\sqrt{\frac{2}{3}}J^{D^{*+}}_{\mu}(x)J^{\Sigma_c^+}(x)+\frac{1}{\sqrt{3}}J^{D^{*0}}_{\mu}(x)J^{\Sigma_c^{++}}(x)\, ,\nonumber\\
J_{\frac{1}{2};\mu\nu}^{D^{*}\Sigma_c^*}(x)&=&\frac{1}{\sqrt{3}}J^{D^{*+}}_{\mu}(x)J^{\Sigma_c^{*+}}_{\nu}(x)-\sqrt{\frac{2}{3}}J^{D^{*0}}_{\mu}(x)J^{\Sigma_c^{*++}}_{\nu}(x)+(\mu\leftrightarrow\nu)\, ,\nonumber\\
J_{\frac{3}{2};\mu\nu}^{D^{*}\Sigma_c^*}(x)&=&\sqrt{\frac{2}{3}}J^{D^{*+}}_{\mu}(x)J^{\Sigma_c^{*+}}_{\nu}(x)+\frac{1}{\sqrt{3}}J^{D^{*0}}_{\mu}(x)J^{\Sigma_c^{*++}}_{\nu}(x)+(\mu\leftrightarrow\nu)\, ,
\end{eqnarray}

\begin{eqnarray}
J_{\frac{1}{2}}^{B\Sigma_b}(x)&=&\frac{1}{\sqrt{3}}J^{B^0}(x)J^{\Sigma_b^0}(x)-\sqrt{\frac{2}{3}}J^{B^-}(x)J^{\Sigma_b^{+}}(x) \, , \nonumber\\
J_{\frac{3}{2}}^{B\Sigma_b}(x)&=&\sqrt{\frac{2}{3}}J^{B^0}(x)J^{\Sigma_b^0}(x)+\frac{1}{\sqrt{3}}J^{B^-}(x)J^{\Sigma_b^{+}}(x)\, ,\nonumber\\
J_{\frac{1}{2};\mu}^{B\Sigma_b^*}(x)&=&\frac{1}{\sqrt{3}}J^{B^0}(x)J^{\Sigma_b^{*0}}_{\mu}(x)-\sqrt{\frac{2}{3}}J^{B^-}(x)J^{\Sigma_b^{*+}}_{\mu}(x)\, ,\nonumber\\
J_{\frac{3}{2};\mu}^{B\Sigma_b^*}(x)&=&\sqrt{\frac{2}{3}}J^{B^0}(x)J^{\Sigma_b^{*0}}_{\mu}(x)+\frac{1}{\sqrt{3}}J^{B^-}(x)J^{\Sigma_b^{*+}}_{\mu}(x)\, ,\nonumber\\
J_{\frac{1}{2};\mu}^{B^{*}\Sigma_b}(x)&=&\frac{1}{\sqrt{3}}J^{B^{*0}}_{\mu}(x)J^{\Sigma_b^0}(x)-\sqrt{\frac{2}{3}}J^{B^{*-}}_{\mu}(x)J^{\Sigma_b^{+}}(x)\, ,\nonumber\\
J_{\frac{3}{2};\mu}^{B^{*}\Sigma_b}(x)&=&=\sqrt{\frac{2}{3}}J^{B^{*0}}_{\mu}(x)J^{\Sigma_b^0}(x)+\frac{1}{\sqrt{3}}J^{B^{*-}}_{\mu}(x)J^{\Sigma_b^{+}}(x)\, ,\nonumber\\
J_{\frac{1}{2};\mu\nu}^{B^{*}\Sigma_b^*}(x)&=&\frac{1}{\sqrt{3}}J^{B^{*0}}_{\mu}(x)J^{\Sigma_b^{*0}}_{\nu}(x)-\sqrt{\frac{2}{3}}J^{B^{*-}}_{\mu}(x)J^{\Sigma_b^{*+}}_{\nu}(x)+(\mu\leftrightarrow\nu)\, ,\nonumber\\
J_{\frac{3}{2};\mu\nu}^{B^{*}\Sigma_b^*}(x)&=&\sqrt{\frac{2}{3}}J^{B^{*0}}_{\mu}(x)J^{\Sigma_b^{*0}}_{\nu}(x)+\frac{1}{\sqrt{3}}J^{B^{*-}}_{\mu}(x)J^{\Sigma_b^{*+}}_{\nu}(x)+(\mu\leftrightarrow\nu)\, ,
\end{eqnarray}
where the subscripts $\frac{1}{2}$ and $\frac{3}{2}$ represent the isospin quantum numbers $I$, moreover, the currents $J_{\frac{1}{2}}(x)$ and $J_{\frac{3}{2}}(x)$ are the isospin eigenstates $|II_3\rangle=|\frac{1}{2}\frac{1}{2}\rangle$ and $|II_3\rangle=|\frac{3}{2}\frac{1}{2}\rangle$, respectively. Based on the above currents, the two-point correlation functions $\Pi(p)$, $\Pi_{\mu\nu}(p)$ and $\Pi_{\mu\nu\alpha\beta}(p)$ are written as,
\begin{eqnarray}
 \Pi(p)&=&\textsf{i}\int d^4x e^{\textsf{i}p\cdot x}\langle 0 |\mathcal{T}\left\{ J (x) \bar{J}(0) \right\}| 0\rangle \, ,\nonumber\\
\Pi_{\mu\nu}(p)&=&\textsf{i}\int d^4x e^{\textsf{i}p\cdot x}\langle 0 |\mathcal{T}\left\{ J_{\mu} (x) \bar{J}_{\nu}(0) \right\}| 0\rangle \, ,\nonumber\\
\Pi_{\mu\nu\alpha\beta}(p)&=&\textsf{i}\int d^4x e^{\textsf{i}p\cdot x}\langle 0 |\mathcal{T}\left\{ J_{\mu\nu} (x) \bar{J}_{\alpha\beta}(0) \right\}| 0\rangle \, ,
\end{eqnarray}
where the $\mathcal{T}$ is the time-order operator, the currents
\begin{eqnarray}
J(x)&=&J_{\frac{1}{2}}^{D\Sigma_c}(x)\, ,\,\,\, J_{\frac{3}{2}}^{D\Sigma_c}(x)\, ,\,\,\,J_{\frac{1}{2}}^{B\Sigma_b}(x)\, , \,\,\, J_{\frac{3}{2}}^{B\Sigma_b}(x)\, ,\nonumber\\
J_{\mu} (x)&=&J_{\frac{1}{2};\mu}^{D\Sigma_c^*}(x)\, ,\,\,\,J_{\frac{3}{2};\mu}^{D\Sigma_c^*}(x)\, ,\,\,\,J_{\frac{1}{2};\mu}^{D^{*}\Sigma_c}(x)\,,\,\,\,J_{\frac{3}{2};\mu}^{D^{*}\Sigma_c}(x)\, ,\nonumber\\
&&J_{\frac{1}{2};\mu}^{B\Sigma_b^*}(x)\, ,\,\,\,J_{\frac{3}{2};\mu}^{B\Sigma_b^*}(x)\, ,\,\,\,J_{\frac{1}{2};\mu}^{B^{*}\Sigma_b}(x)\,,\,\,\,J_{\frac{3}{2};\mu}^{B^{*}\Sigma_b}(x)\, , \nonumber\\
 J_{\mu\nu} (x)&=&J_{\frac{1}{2};\mu\nu}^{D^{*}\Sigma_c^*}(x)\, ,\,\,\,J_{\frac{3}{2};\mu\nu}^{D^{*}\Sigma_c^*}(x)\, ,\,\,\,J_{\frac{1}{2};\mu\nu}^{B^{*}\Sigma_b^*}(x)\, ,\,\,\,J_{\frac{3}{2};\mu\nu}^{B^{*}\Sigma_b^*}(x)\, ,
  \end{eqnarray}
 which have the spin-parity $J^P=\frac{1}{2}^-$, $\frac{3}{2}^-$ and $\frac{5}{2}^-$, respectively. It is worth mentioning that the negative-parity currents could  couple potentially with the states having positive or negative parity \cite{WZG-penta-cc-EPJC-70,WZG-penta-cc-IJMPA-2050003,WZG-penta-cc-EPJC-43,
 WZG-penta-cc-IJMPA-2150071,WZG-penta-cc-NPB-163}. In the framework of QCD sum rules, a complete set of intermediate hadronic states with both positive-parity and negative-parity are inserted into the correlation functions $\Pi(p)$, $\Pi_{\mu\nu}(p)$ and $\Pi_{\mu\nu\alpha\beta}(p)$, then, the contributions of the ground states are separated. Thus, the correlation functions $\Pi(p)$, $\Pi_{\mu\nu}(p)$ and $\Pi_{\mu\nu\alpha\beta}(p)$ at the hadronic sides appear  as,
\begin{eqnarray}\label{CF-HS-1}
  \Pi(p) & = & {\lambda^{-}_{\frac{1}{2}}}^2  {\!\not\!{p}+ M_{-} \over M_{-}^{2}-p^{2}  } +  {\lambda^{+}_{\frac{1}{2}}}^2  {\!\not\!{p}- M_{+} \over M_{+}^{2}-p^{2}  } +\cdots  \, ,\nonumber\\
  &=&\Pi_{\frac{1}{2}}^1(p^2)\!\not\!{p}+\Pi_{\frac{1}{2}}^0(p^2)\, ,
\end{eqnarray}
 \begin{eqnarray}\label{CF-HS-2}
   \Pi_{\mu\nu}(p) & = & {\lambda^{-}_{\frac{3}{2}}}^2  {\!\not\!{p}+ M_{-} \over M_{-}^{2}-p^{2}  } \left(- g_{\mu\nu}\right)+  {\lambda^{+}_{\frac{3}{2}}}^2  {\!\not\!{p}- M_{+} \over M_{+}^{2}-p^{2}  } \left(- g_{\mu\nu}\right)   +\cdots  \, ,\nonumber\\
   &=&-\Pi_{\frac{3}{2}}^1(p^2)\!\not\!{p}\,g_{\mu\nu}-\Pi_{\frac{3}{2}}^0(p^2)\,g_{\mu\nu}+\cdots\, ,
\end{eqnarray}
 \begin{eqnarray}\label{CF-HS-3}
\Pi_{\mu\nu\alpha\beta}(p) & = & {\lambda^{-}_{\frac{5}{2}}}^2  {\!\not\!{p}+ M_{-} \over M_{-}^{2}-p^{2}  } \left( g_{\mu\alpha}g_{\nu\beta}+g_{\mu\beta}g_{\nu\alpha}
\right)+ {\lambda^{+}_{\frac{5}{2}}}^2  {\!\not\!{p}- M_{+} \over M_{+}^{2}-p^{2}  } \left( g_{\mu\alpha}g_{\nu\beta}+g_{\mu\beta}g_{\nu\alpha}\right) +\cdots \, , \nonumber\\
&=&\Pi_{\frac{5}{2}}^1(p^2)\!\not\!{p}\left( g_{\mu\alpha}g_{\nu\beta}+g_{\mu\beta}g_{\nu\alpha}\right)+\Pi_{\frac{5}{2}}^0(p^2)\,\left( g_{\mu\alpha}g_{\nu\beta}+g_{\mu\beta}g_{\nu\alpha}\right)+ \cdots \, ,
\end{eqnarray}
where the components $\Pi_{\frac{1}{2}}^1(p^2)$, $\Pi_{\frac{1}{2}}^0(p^2)$, $\Pi_{\frac{3}{2}}^1(p^2)$, $\Pi_{\frac{3}{2}}^0(p^2)$, $\Pi_{\frac{5}{2}}^1(p^2)$ and $\Pi_{\frac{5}{2}}^0(p^2)$ are picked out to study the pentaquark molecular states,  the subscripts $\frac{1}{2}$, $\frac{3}{2}$ and $\frac{5}{2}$ stand for the angular momentum. For more technical details, the interested readers can consult Refs.\cite{WZG-penta-cc-EPJC-70,WZG-penta-cc-IJMPA-2050003}. The $\lambda_{\frac{1}{2}/\frac{3}{2}/\frac{5}{2}}^{\mp}$ are the pole residues which embody the couplings  between the currents and molecular states, which are defined as,
\begin{eqnarray}\label{J-lamda-1}
\langle 0| J (0)|\mathcal{P}_{\frac{1}{2}}^{-}(p)\rangle &=&\lambda^{-}_{\frac{1}{2}} U^{-}(p,s) \, ,\nonumber  \\
\langle 0| J (0)|\mathcal{P}_{\frac{1}{2}}^{+}(p)\rangle &=&\lambda^{+}_{\frac{1}{2}}\textsf{i}\gamma_5 U^{+}(p,s) \, ,
 \end{eqnarray}
\begin{eqnarray}\label{J-lamda-2}
\langle 0| J_{\mu} (0)|\mathcal{P}_{\frac{3}{2}}^{-}(p)\rangle &=&\lambda^{-}_{\frac{3}{2}} U^{-}_\mu(p,s) \, , \nonumber \\
\langle 0| J_{\mu} (0)|\mathcal{P}_{\frac{3}{2}}^{+}(p)\rangle &=&\lambda^{+}_{\frac{3}{2}}\textsf{i}\gamma_5 U^{+}_{\mu}(p,s) \, ,
\end{eqnarray}
 \begin{eqnarray}\label{J-lamda-3}
\langle 0| J_{\mu\nu} (0)|\mathcal{P}_{\frac{5}{2}}^{-}(p)\rangle &=&\sqrt{2}\lambda^{-}_{\frac{5}{2}} U^{-}_{\mu\nu}(p,s) \, , \nonumber\\
\langle 0| J_{\mu\nu} (0)|\mathcal{P}_{\frac{5}{2}}^{+}(p)\rangle &=&\sqrt{2}\lambda^{+}_{\frac{5}{2}}\textsf{i}\gamma_5 U^{+}_{\mu\nu}(p,s) \, ,
\end{eqnarray}
 where the $\mathcal{P}_{J}^{\mp}(p)$ represent the ground states with the spin-parity $J^P=J^{\mp}$, the $U^{\pm}(p,s)$ are the Dirac spinors, the $U^{\pm}_\mu(p,s)$ and $U^{\pm}_{\mu\nu}(p,s)$ are the Rarita-Schwinger spinors. For simplicity, the isospin quantum numbers are neglected in Eqs.\eqref{CF-HS-1}-\eqref{J-lamda-3}.

At the QCD sides, the Wick's theorem is applied to contract the quark fields. It is interesting to see that  the correlation function $\Pi^{\frac{3}{2}}(p)$ with the higher isospin is equal to  $\frac{4}{5}$ times the corresponding correlation function  $\Pi^{\frac{1}{2}}(p)$ with the lower isospin, where we add the superscripts $\frac{1}{2}$ and $\frac{3}{2}$ to denote the isospins. Such a relation shows that, in the framework of QCD sum rules, the mass of the higher isospin state is degenerated with the lower one for the states considered in the present work, while for the pole residues, $\lambda^{\frac{3}{2}}=\frac{2}{\sqrt{5}}\lambda^{\frac{1}{2}}$, again the superscripts $\frac{3}{2}$ and $\frac{1}{2}$ denote  the isospins.

After the Wick's contractions, the correlation functions are analytically expressed in terms of full-quark propagators at the QCD sides, the operator product expansion (OPE) are performed  and the full-quark propagators are written as,
\begin{eqnarray}
\notag\ L^{ij}(x)&=& \frac{\textsf{i}x\!\!\!/\delta^{ij}}{2\pi^{2}x^{4}}-\frac{\delta^{ij}}{12}\langle\overline{q}q\rangle-\frac{\delta^{ij}x^2}{192}\langle\overline{q}g_s\sigma G q\rangle-\frac{\textsf{i}\delta^{ij}x^2x\!\!\!/g_s^2\langle\overline{q}q\rangle^2}{7776}  \nonumber \\
&&-\left(t^n\right)^{ij}\left(x\!\!\!/\sigma^{\alpha\beta}+\sigma^{\alpha\beta}x\!\!\!/\right)\frac{\textsf{i}}{32\pi^2x^2}g_s G_{\alpha\beta}^n  \nonumber \\
\notag\ &&-\frac{\delta^{ij}x^4\langle\overline{q}q\rangle\langle GG \rangle}{27648}-\frac{1}{8}\langle\overline{q}^j\sigma^{\alpha\beta}q^i\rangle\sigma_{\alpha\beta}-\frac{1}{4}\langle\overline{q}^j\gamma_\mu q^i\rangle\gamma^\mu+\cdot\cdot\cdot\, ,
\end{eqnarray}

\begin{eqnarray}
\notag\
H_{ij}(x)&=&\frac{\textsf{i}}{(2\pi)^{4}}\int d^{4}ke^{-\textsf{i}k\cdot x}\bigg\{\frac{\delta_{ij}}{k\!\!\!/-m_{H}}-\frac{g_{s}G_{\alpha\beta }^{h}t_{ij}^{h}}{4}\frac{\sigma^{\alpha\beta}(k\!\!\!/+m_{H})+(k\!\!\!/+m_{H})\sigma ^{\alpha
\beta }}{(k^{2}-m_{H}^{2})^{2}}\\
\notag\
&&+\frac{g_{s}D_{\alpha}G_{\beta\mu}^{h}t_{ij}^{h}\left(f^{\mu\beta\alpha}+f^{\mu\alpha\beta}\right)}{3(k^{2}-m_{H}^{2})^{4}}
+\cdot\cdot\cdot\bigg \}\, ,
\end{eqnarray}
\begin{eqnarray}
f^{\mu \alpha \beta }&=&(k\!\!\!/+m_{H})\gamma ^{\mu
}(k\!\!\!/+m_{H})\gamma ^{\alpha }(k\!\!\!/+m_{H})\gamma ^{\beta
}(k\!\!\!/+m_{H})\, ,
\end{eqnarray}
 where the $L^{ij}(x)$ and $H_{ij}(x)$ are the light and heavy full-quark propagators, $t^n=\frac{\lambda^n}{2}$, the $\lambda^n$ are the Gell-Mann matrixes ($n=1,2,\cdot\cdot\cdot,8$), $D_\alpha=\partial_\alpha-ig_sG_\alpha^ht^h$, the $m_H$ represents the mass of the charm quark ($m_c$) or bottom quark ($m_b$), the masses of the light quarks are neglected. At the QCD sides, all the vacuum condensates are carefully analyzed in Ref.\cite{wangxiuwu}. In the present work, the terms $\langle\bar{q}q\rangle$, $\langle\frac{\alpha_s}{\pi}GG\rangle$, $\langle\overline{q}g_s\sigma Gq\rangle$, $\langle\overline{q}q\rangle^2$, $\langle\frac{\alpha_s}{\pi}GG\rangle\langle\bar{q}q\rangle$, $\langle\overline{q}g_s\sigma Gq\rangle\langle\overline{q}q\rangle$, $\langle\bar{q}q\rangle^3$, $\langle\overline{q}g_s\sigma Gq\rangle^2$, $\langle\frac{\alpha_s}{\pi}GG\rangle\langle\overline{q}q\rangle^2$, $\langle\overline{q}g_s\sigma Gq\rangle\langle\bar{q}q\rangle^2$, $\langle\overline{q}q\rangle^4$, $\langle\overline{q}g_s\sigma Gq\rangle^2\langle\overline{q}q\rangle$ and $\langle\frac{\alpha_s}{\pi}GG\rangle\langle\overline{q}q\rangle^3$ are chosen in the operator product expansions, and other tiny contributions are neglected. To determine the spectral densities at the QCD sides, for the light quark propagators, the integrals are solved in the coordinate space, as for the heavy ones, the calculations are performed in the momentum space.

  Considering  the quark-hadron duality, the weight functions $\sqrt{s}\exp\left(-\frac{s}{T^2}\right)$ and $\exp\left(-\frac{s}{T^2}\right)$ are applied to derive the QCD sum rules for the pentaquark molecular states,
\begin{eqnarray}\label{QCDN}
2M_{J}^-{\lambda^{-}_{J}}^2\exp\left( -\frac{{M_{J}^-}^2}{T^2}\right)
&=& \int_{4m_H^2}^{s_0}ds \left[\sqrt{s}\rho^1_{J,QCD}(s)+\rho^0_{J,QCD}(s)\right]\exp\left( -\frac{s}{T^2}\right)\, ,
\end{eqnarray}
\begin{eqnarray}\label{QCDSR-M}
 {M_{J}^-}^2 &=& \frac{-\frac{d}{d \tau}\int_{4m_H^2}^{s_0}ds \,\left[\sqrt{s}\,\rho^1_{J,QCD}(s)+\,\rho^0_{J,QCD}(s)\right]\exp\left(- \tau s\right)}{\int_{4m_H^2}^{s_0}ds \left[\sqrt{s}\,\rho_{J,QCD}^1(s)+\,\rho^0_{J,QCD}(s)\right]\exp\left( -\tau s\right)}\, ,
 \end{eqnarray}
where the $T^2$ is the Borel parameter, $\tau=\frac{1}{T^2}$, the $s_0$ is the continuum threshold parameter of the corresponding  molecular state, the $\rho_{J,QCD}^1(s)$ and $\rho_{J,QCD}^0(s)$ are the QCD spectral densities  from the correlation functions $\Pi_{J}^1(p^2)$ and $\Pi_{J}^0(p^2)$, respectively.

\section{Numerical results and discussions}
For the QCD sum rules, the masses and pole residues are extracted in the Borel platforms based on the numerical calculations. The vacuum condensates are taken as input parameters, in details, at the energy scale $\mu=1\,{\rm GeV}$, the standard values of the vacuum condensates $\langle\overline{q}q\rangle=-(0.24\pm0.01\,{\rm GeV})^3$, $\langle\overline{q}g_s\sigma Gq\rangle=m_0^2\langle\overline{q}q\rangle$, $m_0^2=(0.8\pm0.1)\,{\rm GeV}^2$, $\langle\frac{\alpha_s}{\pi}GG\rangle=(0.33\,{\rm GeV})^4$ \cite{SVZ1,SVZ2,Reinders,ColangeloReview}. The $
\overline{MS}$ masses $m_c(m_c)=1.275\pm0.025\,{\rm GeV}$ and  $m_b(m_b)=4.18\pm0.03\,{\rm GeV}$ are taken from the Particle Data Group \cite{PDG}. Moreover, those input parameters depend on the energy scale,
\begin{eqnarray}
\notag \langle\overline{q}q\rangle(\mu)&=&\langle\overline{q}q\rangle(1{\rm GeV})\left[\frac{\alpha_s(1{\rm GeV})}{\alpha_s(\mu)}\right]^{\frac{12}{33-2n_f}}\, ,\\
\notag \langle\overline{q}g_s\sigma Gq\rangle(\mu)& =&\langle\overline{q}g_s\sigma Gq\rangle(1{\rm GeV})\left[\frac{\alpha_s(1{\rm GeV})}{\alpha_s(\mu)}\right]^{\frac{2}{33-2n_f}}\, ,\\
\notag  m_c(\mu)&=&m_c(m_c)\left[\frac{\alpha_s(\mu)}{\alpha_s(m_c)}\right]^{\frac{12}{33-2n_f}}\, ,\\
\notag  m_b(\mu)&=&m_b(m_b)\left[\frac{\alpha_s(\mu)}{\alpha_s(m_b)}\right]^{\frac{12}{33-2n_f}}\, ,\\
\notag \alpha_s(\mu)&=&\frac{1}{b_0t}\left[1-\frac{b_1}{b_0^2}\frac{\rm{log}\emph{t}}{t}+\frac{b_1^2(\rm{log}^2\emph{t}-\rm{log}\emph {t}-1)+\emph{b}_0\emph{b}_2}{b_0^4t^2}\right]\, ,
\end{eqnarray}
where $t=\rm{log}\frac{\mu^2}{\Lambda_{\emph{QCD}}^2}$, $\emph b_0=\frac{33-2\emph{n}_\emph{f}}{12\pi}$, $b_1=\frac{153-19n_f}{24\pi^2}$, $b_2=\frac{2857-\frac{5033}{9}n_f+\frac{325}{27}n_f^2}{128\pi^3}$ \cite{PDG,Narison}, the $n_f$ is the flavor number, it is set as $n_f=4$ and $n_f=5$ for the doubly-charm and doubly-bottom pentaquark molecular states, respectively,  the corresponding $\Lambda_{QCD}$ which depends on $n_f$ are $296\, \rm{MeV}$ and $213\, \rm{MeV}$, respectively. The energy scale formula is applied to determine the best energy scales of the QCD spectral densities \cite{WZG-penta-mole-CPC,Wang-tetraquark-QCDSR-1,Wang-molecule-QCDSR-1,XWWang-dibaryon-sigma},
\begin{eqnarray}
\mu&=&\sqrt{M_{X/Y/Z/P}^2-4\mathbb{M}_H^2}\, ,
\end{eqnarray}
where the $M_{X/Y/Z/P}$ represents the mass of an exotic state. For the charmed pentaquark molecular states, the  effective charm quark mass, $\mathbb{M}_c=1.85\pm0.01$ \rm GeV \cite{XWWang-penta-mole,WZG-penta-mole-CPC}. For the bottom pentaquark molecular states, the  effective bottom quark mass, $\mathbb{M}_b=5.17\pm0.03$ \rm GeV \cite{WZG-mb-value1,WZG-mb-value2,WZG-mb-value3}.

The pole dominance and convergence of the operator product expansion are two basic requirements of the QCD sum rules. The two criteria are studied quantitatively  via the pole contributions (PC) and dimensional contributions, respectively,
\begin{eqnarray}
{\rm PC}&=&\frac{\int_{4m_H^2}^{s_0}ds\left[\sqrt{s}\rho_{QCD}^1(s)+\rho_{QCD}^0(s)\right]\exp\left(-\frac{s}{T^2}\right)}
{\int_{4m_H^2}^{\infty}ds\left[\sqrt{s}\rho_{QCD}^1(s)+\rho_{QCD}^0(s)\right]\exp\left(-\frac{s}{T^2}\right)}\, ,
\end{eqnarray}
\begin{eqnarray}
d(n)&=&\frac{\int_{4m_H^2}^{s_0}ds\left[\sqrt{s}\rho_{QCD;n}^1(s)+\rho_{QCD;n}^0(s)\right]\exp\left(-\frac{s}{T^2}\right)}
{\int_{4m_H^2}^{s_0}ds\left[\sqrt{s}\rho_{QCD}^1(s)+\rho_{QCD}^0(s)\right]\exp\left(-\frac{s}{T^2}\right)}\, ,
\end{eqnarray}
where other  subscripts or superscripts used to mark the particular states are neglected for simplicity, the $\rho_{QCD;n}^1(s)$ and $\rho_{QCD;n}^0(s)$ are the spectral densities with the vacuum condensates of dimension $n$ from the total spectral densities $\rho_{QCD}^1(s)$ and $\rho_{QCD}^0(s)$, respectively. The normalization of the dimensional contribution $d(n)$ is usually studied via the definition  $D(n)=\frac{|d(n)|}{\sum\limits_n|d(n)|}$ to make the convergent behavior  of the OPE. The Borel platforms are determined via trial and error, the essential preconditions are  the above mentioned two basic criteria, furthermore, the extracted masses satisfy the energy scale formula and the continuum threshold parameters $s_0$ satisfy the relations $\sqrt{s_0}=m_{\mathcal{P}_c}+0.6\sim 0.7\,{\rm{GeV}}$, $\sqrt{s_0}=m_{\mathcal{P}_b}+0.8\sim 0.9\,{\rm{GeV}}$, where the $m_{\mathcal{P}_c}$ and $m_{\mathcal{P}_b}$ are the masses of the doubly-charm and doubly-bottom pentaquark molecular states, respectively \cite{XWWang-penta-mole,WZG-penta-mole-CPC,WZGWX-penta-mole}.

The uncertainties of the extracted masses and pole residues $\Delta f$ are approximately estimated via,
\begin{eqnarray}
\Delta f=\sqrt{\sum\limits_i\left[f(\bar x_i\pm \Delta x_i)-f(\bar x_i)\right]^2}\, ,
\end{eqnarray}
where the $f(x)$ stands for the analytical expression of the mass or the pole residue with respect to the input parameters $x$, the $\bar x_i$ stands for the central values of parameters $\langle\bar qq\rangle$, $\langle\overline{q}g_s\sigma Gq\rangle$, $\cdot\cdot\cdot\,$, while the $\Delta x_i$ are their uncertainties.

\begin{figure}
 \centering
 \includegraphics[totalheight=5cm,width=7cm]{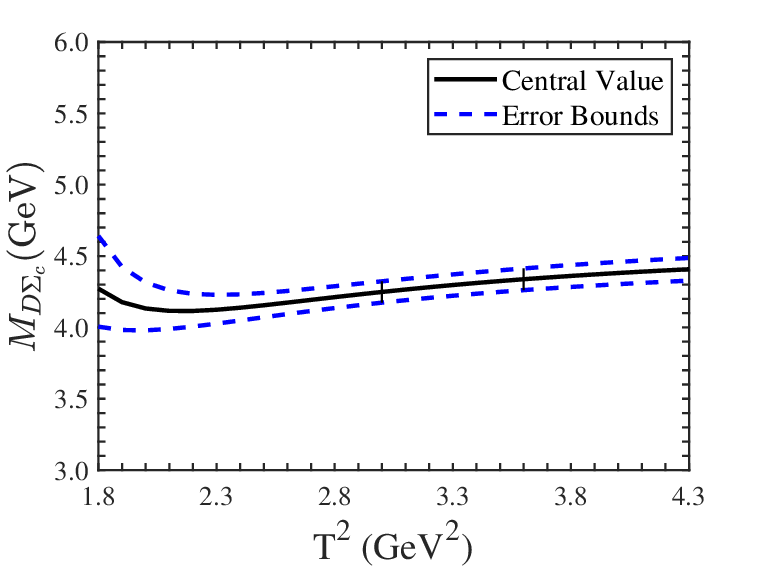}
 \includegraphics[totalheight=5cm,width=7cm]{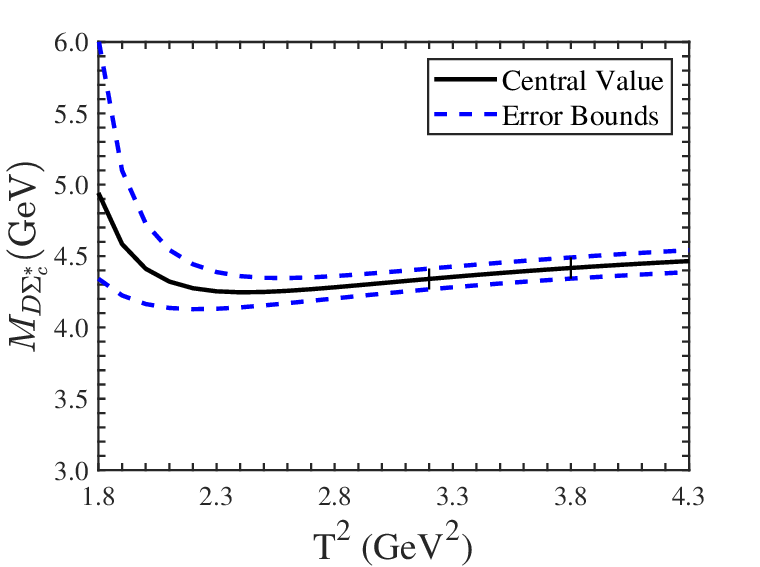}
 \includegraphics[totalheight=5cm,width=7cm]{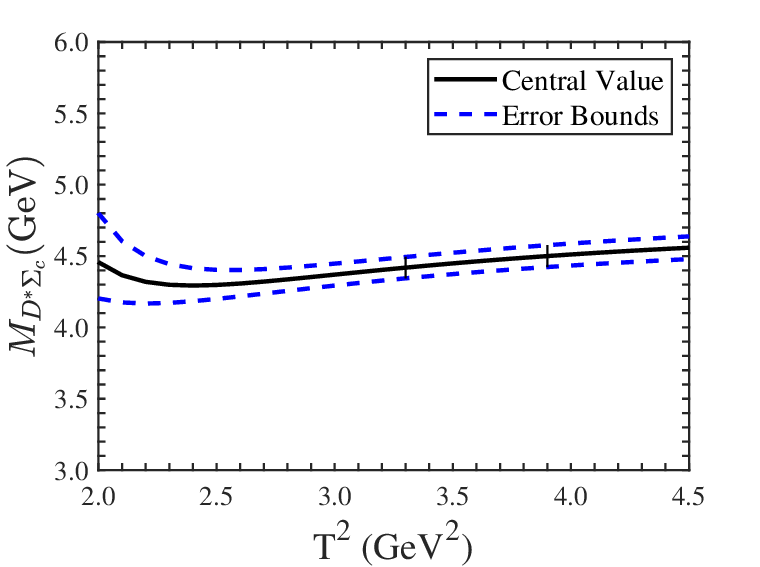}
 \includegraphics[totalheight=5cm,width=7cm]{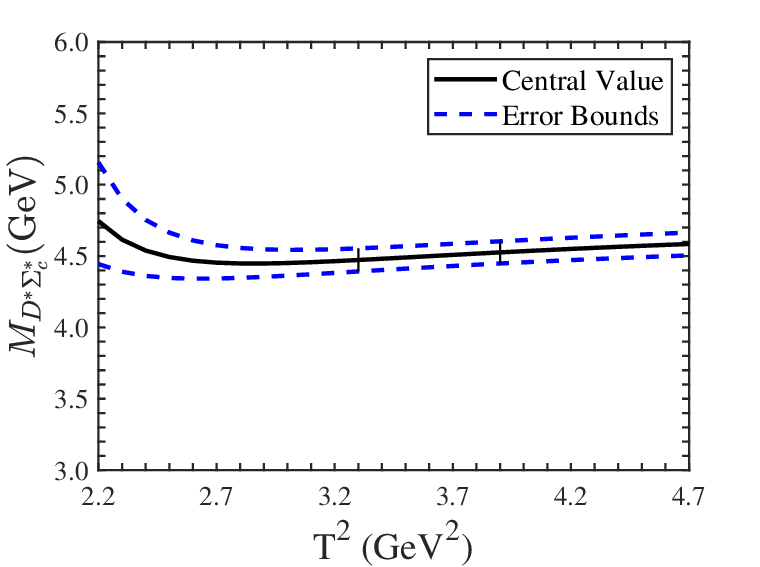}
  \caption{The $M-T^2$ curves of the doubly-charm  pentaquark molecular states.}\label{mass-fig}
\end{figure}

\begin{figure}
 \centering
 \includegraphics[totalheight=5cm,width=7cm]{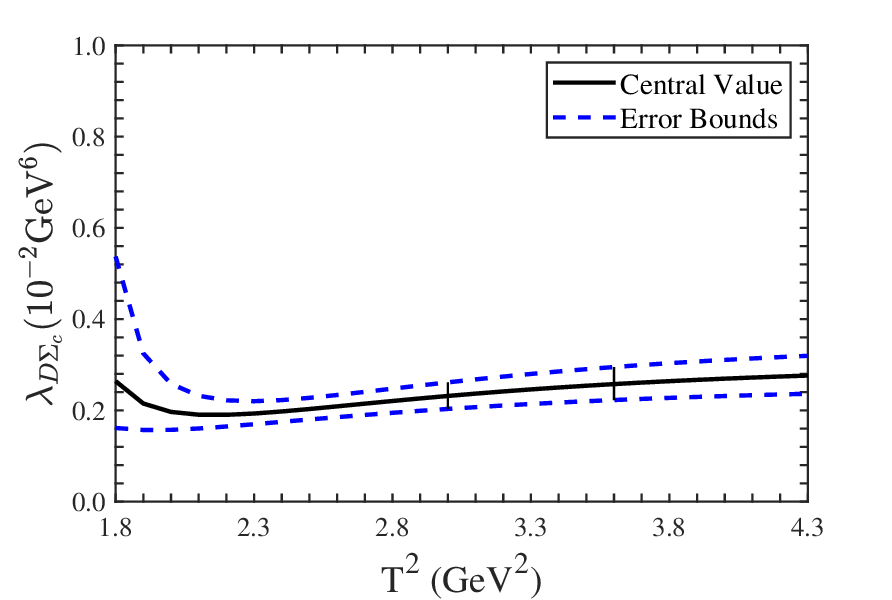}
 \includegraphics[totalheight=5cm,width=7cm]{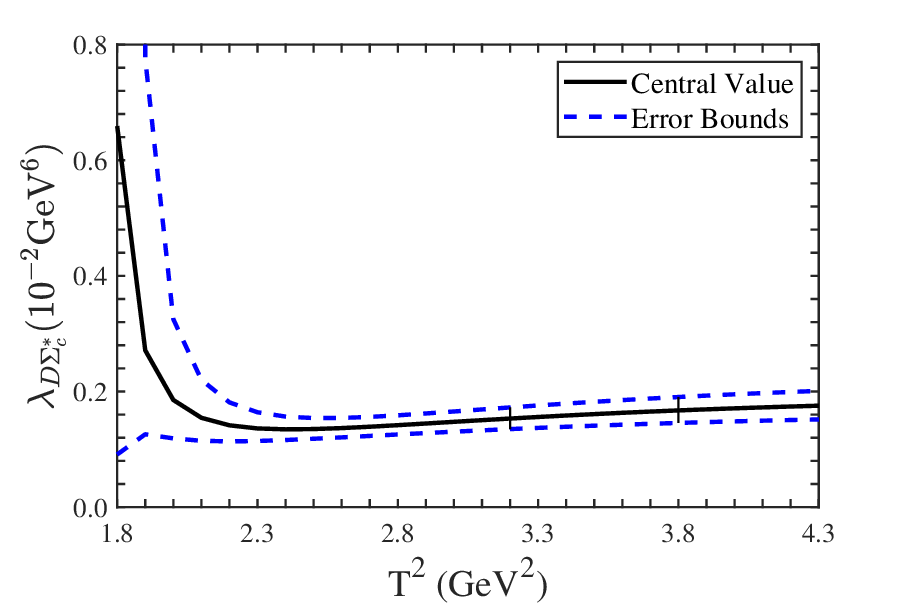}
 \includegraphics[totalheight=5cm,width=7cm]{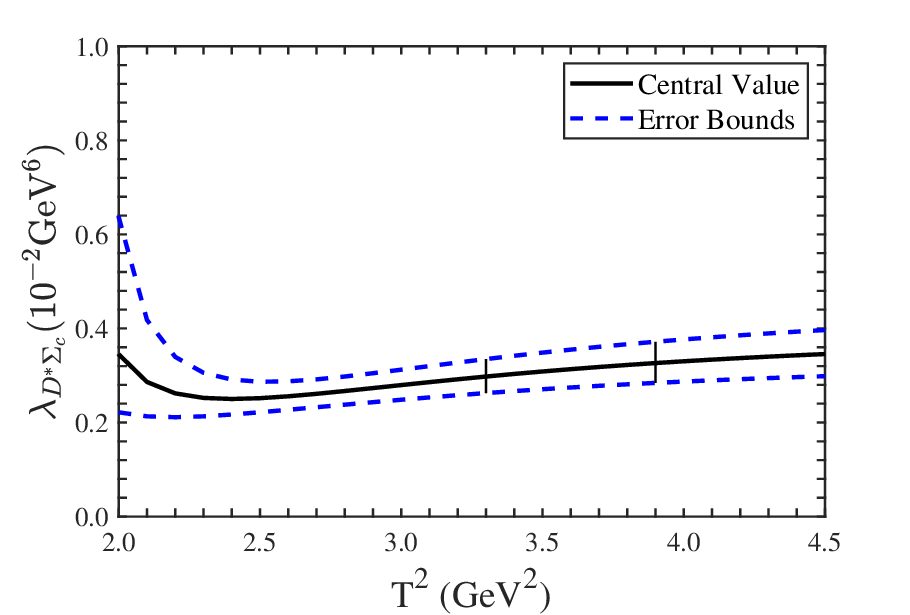}
 \includegraphics[totalheight=5cm,width=7cm]{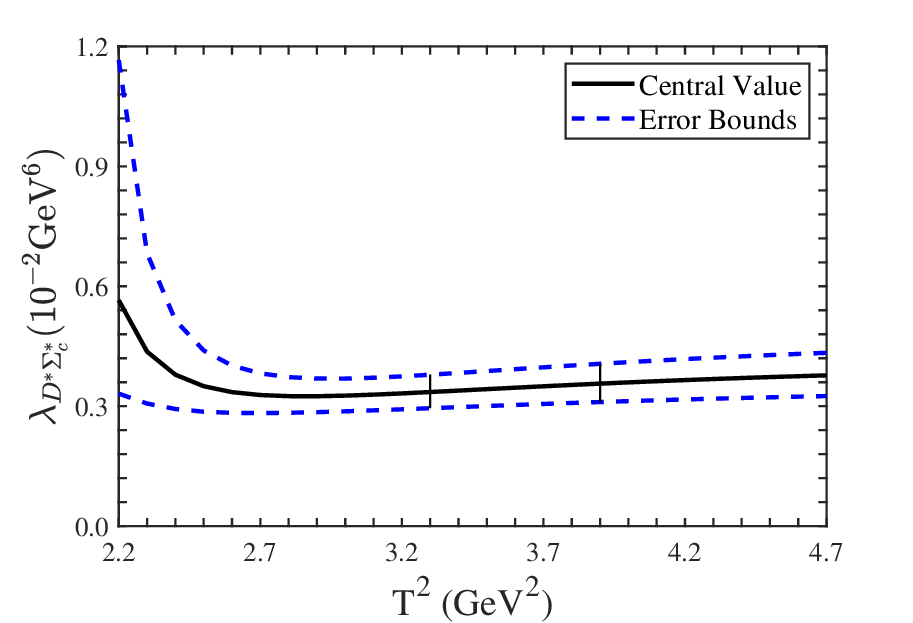}
  \caption{The $\lambda-T^2$ curves of the doubly-charm  pentaquark molecular states}\label{lamt-fig}
\end{figure}

The $M-T^2$ and $\lambda-T^2$ curves are shown in the Fig.\ref{mass-fig} and Fig.\ref{lamt-fig}, respectively, where the regions between the two short lines are the Borel platforms for the doubly-charm pentaquark molecular states. The detailed numerical results extracted from the Borel platforms are given in the Table \ref{BorelP} and Table \ref{mass-residue}. 

\begin{figure}
 \centering
 \includegraphics[totalheight=5cm,width=7cm]{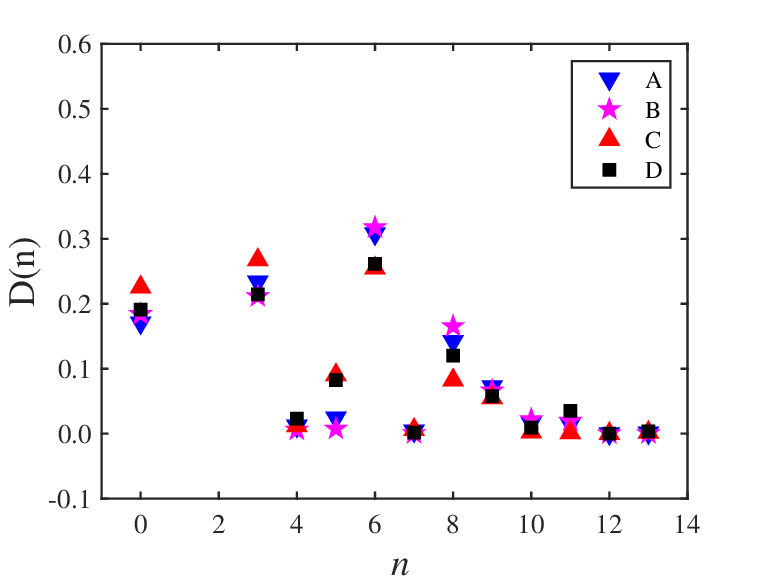}
 \includegraphics[totalheight=5cm,width=7cm]{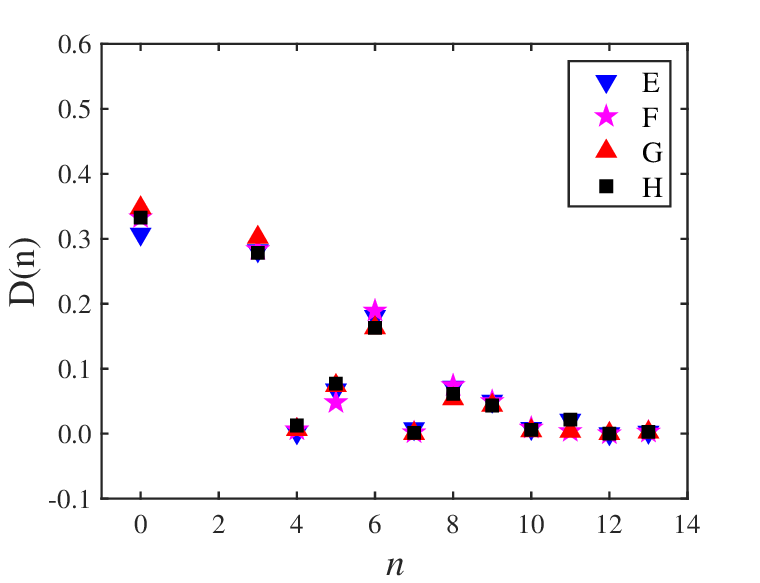}
 \caption{The dimensional contributions of the doubly-charm and doubly-bottom pentaquark molecular states, where the $A$, $B$, $C$, $D$, $E$, $F$, $G$ and $H$ represent the $D\Sigma_c$, $D\Sigma_c^*$, $D^*\Sigma_c$, $D^*\Sigma_c^*$, $B\Sigma_b$, $B\Sigma_b^*$, $B^*\Sigma_b$ and $B^*\Sigma_b^*$  states, respectively.} \label{Dn1-fig}
\end{figure}

For the $D(n)$, the numerical results are determined by the central values of the input parameters, from the Fig.\ref{Dn1-fig}, we can see  that the high dimensional vacuum condensates play minor roles, in details, the $D(12)$ and $D(13)$ are less than $0.01\%$ and $0.37\%$, respectively, thus, the convergency of the OPE holds well for the QCD sum rules in the present work. It also shows that the vacuum condensates  $\langle \bar{q}q\rangle$ and $\langle \bar{q}q\rangle^2$ play the most important roles  for all the doubly-heavy pentaquark molecular states. The other major contributions are mainly from the $\langle\overline{q}g_s\sigma Gq\rangle$ and $\langle \bar{q}q\rangle\langle\overline{q}g_s\sigma Gq\rangle$. As for the $D(4)$, which is due to the condensate $\langle\frac{\alpha_s}{\pi}GG\rangle$, its value is less than $2.31\%$ for all the pentaquark molecular  states.

As shown in the Table \ref{BorelP}, the widths of the Borel platforms are uniformly set as $T_{max}^2-T_{min}^2=0.6\,\rm{GeV}^2$ and $T_{max}^2-T_{min}^2=1.6\,\rm{GeV}^2$ for the doubly-charm and doubly-bottom pentaquark molecular states, respectively. The pole contribution is about $(40-60)\%$ for every state, this indicates that the pole dominance holds well, moreover, the chosen energy scales satisfy  the energy scale formula. 

In the Table \ref{mass-residue}, the thresholds of the $D^{(*)}\Sigma_c^{(*)}$ and $B^{(*)}\Sigma_b^{(*)}$ scattering  states are given based on the masses  from the Particle Data Group \cite{PDG}. Numerical results of the masses of the considered doubly-charm and doubly-bottom pentaquark molecular states are all below the corresponding thresholds for a few dozens of $\rm{MeV}$, if neglecting  their uncertainties. The pole residues of the doubly-charm and doubly-bottom pentaquark molecular states have the orders of magnitude of $10^{-3}\,\rm{GeV^6}$ and $10^{-2}\,\rm{GeV^6}$, respectively.

Taking the conservation of the isospin, angular momentum and parity  into consideration, the possible strong decay channels for the predicted doubly-charm pentaquark molecular  states are $\Xi_{cc}+\pi/\eta/\rho/\omega$.

\begin{table}
\begin{center}
\begin{tabular}{|c|c|c|c|c|c|c|c|c|}\hline\hline
                         &\ $IJ^P$                        &\ $T^2 ({\rm GeV}^2)$   &\ $\sqrt{s_0}({\rm GeV})$    &\ $\mu ({\rm GeV})$   &\ $\rm PC$  \\
\hline
$D\Sigma_c$        &\ $\frac{1}{2}\frac{1}{2}^-$    &\ $3.0-3.6$  &\   $4.95\pm 0.10$  &\ $2.19$   &\ $(41-61)\%$   \\
\hline
$D\Sigma_c^*$      &\ $\frac{1}{2}\frac{3}{2}^-$    &\ $3.2-3.8$  &\   $5.04\pm 0.10$  &\ $2.34$   &\ $(42-60)\% $   \\
\hline
$D^*\Sigma_c$      &\ $\frac{1}{2}\frac{3}{2}^-$    &\ $3.3-3.9$  &\   $5.13\pm 0.10$  &\ $2.49$   &\ $(40-58)\% $  \\
\hline
$D^*\Sigma_c^*$    &\ $\frac{1}{2}\frac{5}{2}^-$    &\ $3.3-3.9$  &\   $5.12\pm0.10$  &\ $2.56$   &\ $(40-57)\% $   \\
\hline\hline

$B\Sigma_b$        &\ $\frac{1}{2}\frac{1}{2}^-$    &\ $8.6-10.2$  &\   $11.93\pm0.10$  &\ $3.95$   &\ $(41-61)\%$   \\
\hline
$B\Sigma_b^*$      &\ $\frac{1}{2}\frac{3}{2}^-$    &\ $8.8-10.4$  &\   $11.98\pm0.10$  &\ $4.01$   &\ $(42-62)\% $   \\
\hline
$B^*\Sigma_b$      &\ $\frac{1}{2}\frac{3}{2}^-$    &\ $9.0-10.6$  &\   $12.00\pm0.10$  &\ $4.09$   &\ $(40-60)\% $  \\
\hline
$B^*\Sigma_b^*$    &\ $\frac{1}{2}\frac{5}{2}^-$    &\ $9.0-10.6$  &\   $12.02\pm0.10$  &\ $4.15$   &\ $(42-61)\% $   \\

\hline\hline
\end{tabular}
\end{center}
\caption{ The Borel parameters, continuum threshold parameters, energy scales and pole contributions. }\label{BorelP}
\end{table}

\begin{table}
\begin{center}
\begin{tabular}{|c|c|c|c|c|c|c|c|c|}\hline\hline
&\ $IJ^P$    &\ $M({\rm GeV})$    &\  $\lambda(10^{-3}{\rm GeV}^6)$     & Meson-Baryon Pairs & Thresholds (MeV)  \\ \hline

$D\Sigma_c$   &\ $\frac{1}{2}\frac{1}{2}^-$  &\  $4.30^{+0.07}_{-0.08}$  &\  $2.46^{+0.34}_{-0.32}$   & $(D^+\Sigma_c^+, D^0\Sigma_c^{++})$ & $(4322, 4319)$\\
\hline
$D\Sigma_c^*$      &\ $\frac{1}{2}\frac{3}{2}^-$      &\  $4.38^{+0.07}_{-0.07}$  &\   $1.61^{+0.21}_{-0.20}$  & $(D^+\Sigma_c^{*+}, D^0\Sigma_c^{*++})$ & $(4387, 4383)$  \\
\hline
$   D^*\Sigma_c$      &\ $\frac{1}{2}\frac{3}{2}^-$      &\  $4.46^{+0.08}_{-0.08}$  &\   $3.13^{+0.41}_{-0.39}$  & $(D^{*+}\Sigma_c^{+}, D^{*0}\Sigma_c^{++})$& $(4463, 4461)$ \\
\hline
$D^*\Sigma_c^*$    & \ $\frac{1}{2}\frac{5}{2}^-$     &\  $4.50^{+0.08}_{-0.08}$  &\   $3.46^{+0.46}_{-0.43}$  & $(D^{*+}\Sigma_c^{*+}, D^{*0}\Sigma_c^{*++})$ & $(4527, 4525)$\\ \hline\hline

$B\Sigma_b$        &\ $\frac{1}{2}\frac{1}{2}^-$      &\  $11.07^{+0.08}_{-0.08}$  &\  $19.89^{+2.65}_{-2.57}$   & $(B^0\Sigma_b^0, B^-\Sigma_b^{+})$&$(11093, 11090)$  \\
\hline
$B\Sigma_b^*$      &\ $\frac{1}{2}\frac{3}{2}^-$      &\  $11.09^{+0.08}_{-0.08}$  &\   $12.10^{+1.56}_{-1.51}$  & $(B^0\Sigma_b^{*0}, B^-\Sigma_b^{*+})$&$(11112, 11110)$  \\
\hline
$   B^*\Sigma_b$      &\ $\frac{1}{2}\frac{3}{2}^-$      &\  $11.12^{+0.08}_{-0.08}$  &\   $22.61^{+2.92}_{-2.84}$  & $(B^{*0}\Sigma_b^{0}, B^{*-}\Sigma_b^{+})$&$(11138, 11135)$ \\
\hline
$B^*\Sigma_b^*$    & \ $\frac{1}{2}\frac{5}{2}^-$     &\  $11.14^{+0.08}_{-0.08}$  &\   $26.37^{+3.34}_{-3.25}$  & $(B^{*0}\Sigma_b^{*0}, B^{*-}\Sigma_b^{*+})$&$(11158, 11155)$ \\
\hline\hline
\end{tabular}
\end{center}
\caption{ The masses, pole residues, meson-baryon thresholds for the doubly-charm and doubly-bottom pentaquark molecular states. The corresponding ones with the isospin $I=\frac{3}{2}$ have the same masses, while the pole residues are re-scaled by a factor $\frac{2}{\sqrt{5}}$.  }\label{mass-residue}
\end{table}

\section{Conclusions}
In the present work, the doubly-heavy  pentaquark molecular states $D^{(*)}\Sigma_c^{(*)}$ and $B^{(*)}\Sigma_b^{(*)}$ are studied via the QCD sum rules in details. The color singlet-singlet type currents with the definite $IJ^P$ are constructed to interpolate  those pentaquark molecular states. Analytical calculations show that the correlation functions with the higher isospin $\frac{3}{2}$ are equal to $\frac{4}{5}$ times that of the lower ones, thus, the masses of the pentaquark molecular states  with the higher isospin $\frac{3}{2}$ are equal to the corresponding ones with the lower isospin $\frac{1}{2}$. Finally, the Borel platforms are determined via trial and error, and the masses and pole residues of the doubly-heavy pentaquark molecular states are extracted from the Borel windows. Numerical results show that the masses of the  pentaquark molecular states are all below the corresponding  meson-baryon thresholds, and they maybe form the bound states, we can search for those pentaquark molecular states in the decay channels $\Xi_{cc}+\pi/\eta/\rho/\omega$ in the future.

\section*{Acknowledgements}
This work is supported by National Natural Science Foundation, Grant Number 12175068 and the Fundamental Research Funds for the Central Universities of China.

\end{document}